\documentstyle[12pt]{article}

\newcommand{\cp}{C\hspace{-1mm}P^1}

\def\bear{\begin{eqnarray}}
\def\enar{\end{eqnarray}}
\def\rf#1{(\ref{eq:#1})}
\def\lab#1{\label{eq:#1}}
\def\nonu{\nonumber}
\def\nn{\nonumber \\}
\def\br{\begin{eqnarray}}
\def\er{\end{eqnarray}}
\def\be{\begin{equation}}
\def\ee{\end{equation}}

\def\({\left(}
\def\){\right)}

\newcommand{\ct}[1]{\cite{#1}}
\newcommand{\bi}[1]{\bibitem{#1}}
\def\lskip{\vskip\baselineskip\vskip-\parskip\noindent}
\relax

\def\l{\lambda}

\def\o{\over}

\def\pa{\partial}

\def\th{\theta}

\def\tp0{\Theta_{+}^{(0)}}
\def\tm0{\Theta_{-}^{(0)}}

\def\u2{\mid u\mid^2}

\def\cj{{\cal J}}

\def\f#1#2#3 {f^{#1#2}_{#3}}

\def\win1{{\sf w_{1+\infty}}}

\def\Win1{{\sf W_{1+\infty}}}

\def\rlx{\relax\leavevmode}
\def\inbar{\vrule height1.5ex width.4pt depth0pt}
\def\IZ{\rlx\hbox{\sf Z\kern-.4em Z}}
\def\IR{\rlx\hbox{\rm I\kern-.18em R}}
\def\IC{\rlx\hbox{\,$\inbar\kern-.3em{\rm C}$}}
\def\IN{\rlx\hbox{\rm I\kern-.18em N}}
\def\IO{\rlx\hbox{\,$\inbar\kern-.3em{\rm O}$}}
\def\IP{\rlx\hbox{\rm I\kern-.18em P}}
\def\IQ{\rlx\hbox{\,$\inbar\kern-.3em{\rm Q}$}}
\def\IF{\rlx\hbox{\rm I\kern-.18em F}}
\def\IG{\rlx\hbox{\,$\inbar\kern-.3em{\rm G}$}}
\def\IH{\rlx\hbox{\rm I\kern-.18em H}}
\def\II{\rlx\hbox{\rm I\kern-.18em I}}
\def\IK{\rlx\hbox{\rm I\kern-.18em K}}
\def\IL{\rlx\hbox{\rm I\kern-.18em L}}
\def\one{\hbox{{1}\kern-.25em\hbox{l}}}
\def\0#1{\relax\ifmmode\mathaccent''7017{#1}%
B        \else\accent23#1\relax\fi}

\def\PRL#1#2#3{{\sl Phys. Rev. Lett.} {\bf#1} (#2) #3}
\def\NPB#1#2#3{{\sl Nucl. Phys.} {\bf B#1} (#2) #3}

\def\PRD#1#2#3{{\sl Phys. Rev.} {\bf D#1} (#2) #3}

\def\PLB#1#2#3{{\sl Phys. Lett.} {\bf B#1} (#2) #3}
\def\JMP#1#2#3{{\sl J. Math. Phys.} {\bf #1} (#2) #3}

\begin{document}
\begin{flushright}
\setcounter{page}{0}

{\large October 2002}\\
\end {flushright}
\vspace*{\fill}
\begin{center}
{\Large \bf Generalized Integrability and the connections between  \\ [0.5cm]
Skyrme-Faddeev  and Yang Mills theories}\\[2.cm]

{\Large Joaqu\'{\i}n S\'anchez-Guill\'en  }\\[5mm]

{Facultad de F\'\i sica, University of Santiago de Compostela \\ 
15706 Santiago de Compostela, Spain}

\end{center}
\begin{abstract}{\small\noindent
Skyrme theory on  $S^2$ (Faddeev coset proposal), is analyzed with a generalization of $0$-curvature integrability,
based on gauge techniques. New expressions valid for models in the sphere are given.
The relation of the minimum energy configurations to gauge vacua is clarified. Consequences of adding a potential term to break
the $SO(3)$ symmetry are discussed.}
\end{abstract}
\vspace*{\fill}
\newpage

Skyrme's original  proposal was to  adjust chiral fields with an stabilizing  quartic derivative term \ct{Sky}. As it is well known, his
new soliton solutions representing nucleons were shown later to correspond to the planar limit  of the non-abelian  gauge theory \ct{Witten}
. Faddeev conjectured a more direct connection to pure QCD, restricting  Skyrme chiral fields to the coset 
$SU(2)/U(1)$ \ct {Fad} and there have been many efforts to reach the Skyrme-Faddev model  from  perturbative functional
integrations of fast frequencies
in special decompositons of the nonabelian gauge fields \ct{gies}.

Non-perturbative progress has generally required recourse to numerical computations  both for ordinary Skyrme \ct{Houg} as well as for Faddeev 
  formulation\ct {Fade}, which was also investigated by lattice methods \ct {vW}. Analytical attempts 
were also initiated \ct {last} in a scheme which used gauge techniques and fields as auxiliary connections to study 
non-linear  systems in higher  dimensions\ct{afg}, generalizing the methods of 2 dimensional field theory. The
 theory will have infinite conserved
 currents {\it if}
its equations of motion, expressed in terms of the auxiliary gauge fields, are independent of the representation of their Lie
 algebra. In higher dimensions
this is exceptional, 
but then   the formalism naturally
gives  constraints for which  sectors of the theory exhibit infinite conserved currents. A zero curvature representation for the Skyrme
Faddeev model was  included in \ct{afz}, among other examples of models defined on a $S^2$ sphere, discussing briefly the corresponding 
integrable sector.

In this Letter we work out the generalized  integrability analysis of the model, having in mind the connection with the
 gauge theory. We pay  special attention to a coincidence found recently \ct{vB} between the static energy and a functional related to that
minimized in  the abelian projections. This  relation of the minimum energy configurations to gauge vacua is clarified and is  an example of 
other possibilities, offered naturally by the generalized  integrablity, of connecting gauge fixings with nonlinear physical systems. Those
can get unexpected simplifications in the analysis, while the former obtain physical interpretations. Another application of the analysis
concerns additional potentials, as suggested by Faddev and Niemi to account for the
$SO(3)$ breaking and avoid global colour problems \ct{fadedual}. Such  explicit breaking of the global symmetry was first suggested in 
\ct {vW} to avoid spontaneous Goldstone modes, incompatbible with the mass gap of pure QCD.

 With  $R^3$ compatified to $S^3$ from 
the finite energy requirement, Faddeev proposal to go to the $SU(2)/U(1)$ 2-sphere, changes the degree by the Hopf map and the
topological charge becomes the linking number of the preimages. The solitons should have then knot configurations  
 and  axial symmetry is the simplest allowed.
The action for the Skyrme-Faddeev model is given by
\be
 S = \int d^4x \(m^2 ( \pa {\bf n}\)^2 )-\( {1 \over e^2}\pa_{\mu} {\bf n} \times \pa_{\nu}{\bf n}\) 
^2 
\lab{sflag}
\ee
where  ${\bf n}$ is a unit field in $SU(2)$ colour space. $m$ is a parameter with  dimension of mass and $e^2$ can be considered
as a strong coupling.
The second
term is the pull back of the area form on
$S^2$, and  it is the tensor used in the Skyrme-Faddeev model to balance the instability of the solitons under
rescaling of the space variables.

A potential term can be added \ct {fadedual} to circumvent the global problems 
with colour in the glueball interpretation. Such  terms, which will be discussed
later, are also required for the pion mass and phenomenological application in the ordinary Skyrme case and for stability in 2+1 dimensions
(the so called baby Skyrme \ct{21}), which is closely related in many ways to the Skyrme Faddeev model . This can be seen in terms of the
complex field
$u$ of the stereographic projection of the $S^2$ defined by ${\bf n}^2 =1$, which is of course  very useful for fields on the sphere, independent of the
space-time dimension,
\be
{\bf n} = {1\o {1+\mid u\mid^2}} \, \( u+u^* , -i \( u-u^* \) , \u2 -1 \) \; ; 
\qquad \quad 
u \equiv u_1 + i u_2 = \frac{n_1 + i n_2}{1 - n_3}
\lab{stereo}
\ee
 The second term is the square of the antisymmetric and
real tensor
\be
h_{\mu\nu}\equiv \frac{-2i}{\( 1+\u2\)^2}\, 
\( \pa_{\mu} u \pa_{\nu} u^* - \pa_{\nu} u \pa_{\mu} u^*\) 
\sim {\bf n}\cdot \( \pa_{\mu} {\bf n} \times \pa_{\nu} {\bf n}\) 
\ee

The  Lagrangian density can be given in a compact form also in terms of $u$

\be
L = {m^2 (\pa u)^2  \over (1+|u|^2)^2} 
-\ {1 \over e^2} {(\pa u \times \pa u^* )^2 \over
(1+|u|^2)^4} 
\ee
and the equations of motion
\bear
0 & = & m^2(\pa_\mu\pa^\mu u - 2 u^* {\pa_\mu u\pa^\mu u\over 1+|u|^2}) -
\nn
&&- {4\over e^2(1+|u|^2)^2} \left \{ 
\pa_\mu\left ( \pa^\mu u^*\pa_\nu u\pa^\nu u - 
\pa^\mu u\pa_\nu u\pa^\nu u^* \right ) \right . \nn
&& \left .
-{2u\over 1+|u|^2}\left ( 
\pa_\mu u\pa^\mu u\pa_\nu u^*\pa^\nu u^* - 
\pa_\mu u \pa^\mu u^* \pa_\nu u\pa^\nu u^*\right ) \right \} \nn
\enar 
 will  become much simplified in our integrability analysis.

 The energy for static configurations of 
the Skyrme-Faddeev model is  given by
\be
E= E_1 + E_2
\ee
with
\br
E_1 &\equiv&  4 m^2 \, \int d^3 x {\mid \nabla u \mid^2 \o { \( 1 + \u2\)^2}} 
\nonu\\
E_2 &\equiv& {8\o e^2}\, \int d^3 x {\( \mid \nabla u \mid^4 
- \( \nabla u \)^2  \( \nabla u^* \)^2 \) \o {\( 1 + \u2\)^4}}
\lab{laener}
\er

Notice  that the above mentioned scaling stability requires, as one immediately sees at first order expanding
in the scaling parameter,
\be
E_1 = E_2
\lab{stab1}
\ee
Models on the sphere in any dimension have a convenient natural formulation in the  approach of \ct{afg}
where  the equations of motion are given by the constant covariance of the dual of an antisymmetric tensor $D_A B=0$,
with respect to a flat gauge connection 
\br
A_\mu &=&
{1\o{\( 1+\mid u \mid^2\) }} \bigg(  \(\pa_{\mu} u + \pa_{\mu} u^* \)\, T_{1}
+i \(\pa_{\mu} u - \pa_{\mu} u^* \)\, T_{2}   \nonu\\
 &+&
i\( u \pa_{\mu} u^* - u^* \pa_{\mu} u \) \, T_3 \bigg).
\lab{lacon}
\er
where $T_{3,\pm}$ are the usual generators of  the $su(2)$ algebra.
That $A_\mu$ is flat follows from the fact that we can write
it in the form $A_\mu = -\pa_\mu g g^{-1}$, where \footnote{This $g$ are the elements which conjugate
Skyrme's {\it hedgehog} Ansatz to the Cartan component and serve to identify its integrable sector \ct{last}.}
$$
g = {1\over\sqrt{1+|u|^2}} 
\left ( \begin{array}{cc} 1&iu\\iu^*&1\end{array}\right ).
$$

For the Skyrme-Faddeev model the (dual of the) 2-form can be taken as
\bear 
B_\mu &=& {1\over 1+|u|^2} \left \{ 
\left (m^2 \pa_\mu u - {4 \over e^2(1+|u|^2)^2}
(\pa_\mu u^*\pa_\nu u\pa^\nu u - \pa_\mu u\pa_\nu u\pa^\nu u^*)
\right ) P_{+1} - \right . \nn
&& \left . 
\left ( m^2\pa_\mu u^* - {4 \over e^2(1+|u|^2)^2}
(\pa_\mu u\pa_\nu u^*\pa^\nu u^* - \pa_\mu u^*\pa_\nu u^*\pa^\nu)
\right ) P_{-1} \right \} \nn
&\equiv& B_\mu^{+1}P_{+1} + B_\mu^{-1} P_{-1}
\enar 
where $P_{\pm 1}$ transform under the spin-1 
representation. A simple calculation shows that 
\bear 
D_\mu B^\mu & = & (\pa^\mu B^{+1}_\mu + A^3_\mu B^{+1}_\mu)P_{+1} +
\nn
&&+ \sqrt{2} (A^\mu_+ B_\mu^{-1} + A^\mu_- B_\mu^{+1}) P_0 + 
(\pa^\mu B^{-1}_\mu - A^3_\mu B^{-1}_\mu)P_{-1} \nn
&=&
 0
\enar 

and we see that the equations of motion are in fact equivalent to the equation
$D_\mu B^\mu=0$ . 
Conjugation of $B$ with $g$ gives conserved currents, which so far are the Noether currents.
But it may be that this geometric representation  contains more information and unravels hidden symmetries
 as in the $2d$ case. 
For that purpose   we can define auxiliary fields $B_\mu^{(j)}$ for any spin $j$
representation in the same way, with $P^{(j)}_m$ trnasforming under the spin -$j$ representation. The
 projection of the result on quantum numbers 0 and
$\pm 1$ is the same as the one given above, but in addition there is now an
extra component of grade $\pm 2$, where for example the projection on
the $+2$ component is proportional to $\pa_\mu u \pa^\mu u$. 
This is completely equivalent to what happens in the usual $\cp$
model, and it leads us to define a restricted model, with infinite conserved currents

\be
J_\mu^{(j)} = g\tilde{B}_\mu^{(j)}g^\dagger \equiv 
\sum_{m=-j}^j J_\mu^{(j),m} \, P^{(j)}_m  \; ; \qquad \qquad 
\mbox{\rm for any positive integer $j$}
\ee

 by imposing the
 constraint 
\footnote{ for the simplest $O(3)$ model in $2+1$ this constraint generalizes the
Cauchy Riemann conditions of the {\it baby Skyrmion} solution.}
\be
( \pa u )^2 =0
\lab{const}
\ee

For this {\it integrable} submodel, the second term in $E_2$ of \rf {laener} vanishes and the stability relation
\rf {stab1} becomes 
\be
\int d^3x \, \cj = \int d^3x \, \cj^2
\lab{stab2}
\ee
where $\cj$ is the dimensionless, positive definite quantity
\be
\cj = {2\o {m^2 e^2}}\, {\mid \nabla u \mid^2 \o { \( 1 + \u2\)^2}}
\lab{jdef}
\ee

The energy for the submodel solutions then becomes
\br
E &=& 2 m^4 e^2 \, \int d^3 x \, \cj \( \cj +1 \) \nonu \\
 &=& 4 m^4 e^2 \, \int d^3 x \, \cj \nonu\\ 
 &=& 4 m^4 e^2 \, \int d^3 x \, \cj^2
\er

Therefore, the submodel solutions  present the spectrum of a rotor, which is the
natural adiabatic step  from a classical geometric picture to the quantum case, as in nuclear
physics or Skyrme model itself. The equation of motion of the submodel
have a useful simple form in terms of  $\cj$
\be
\pa^{\mu }\left (m^2 \( 2 \cj + 1 \) \pa_{\mu} u
\right ) = 0 
\lab{submeq}
\ee
One can easily check \ct {Luiz} that the constraint minimizes the energy and it is compatible  with the stability condition \rf {stab1}
and  the axial symmetry required.
The question is whether there are solutions, field configurations 
minimizing the energy, as we discuss below. 

Of special interest is the application of the geometrical integrability formulation of Skyrme-Faddev (and similar models) to an
intriguing new connection with the gauge theory \ct {vW} and \ct{vB}.
 Performing consecutive ingenious changes of
variables of the adjusted ${\bf n}$ field, the authors found that  its static energy can be written in terms of a flat connection closely related to
the   functional  chosen in the maximal abelian procedure, to fix by minimization to  the
non-abelian theories to an abelian gauge \footnote
{This implements the abelian Higgs phenomenon  and it should correspond
to the monopole condensation scenario of confinement \ct{tH}.}.
In our zero curvature approach, these relations are  present from the beginning, since that flat connection is precisely 
the  one chosen to represent Skyrme-Faddev (and other models on the sphere).
In fact, writing explicitly the components along the step operators of the auxiliary
flat connection \rf{lacon}
(as in eq. (6.58) of \ct{afg}) as

\be
A_j^1 \equiv {\pa_{j} u + \pa_{j} u^* \o{\( 1+\mid u \mid^2\) }}
\qquad \qquad
A_j^2 \equiv i \; {\pa_{j} u - \pa_{j} u^* \o{\( 1+\mid u \mid^2\) }}
\ee
one has
\br
A_j^1 A_j^1 + A_j^2 A_j^2 = 4 \;
\frac{\mid \nabla u \mid^2}{\( 1+\mid u \mid^2\)^2 }
\er
In addition
\be
A_i^1 A_j^2 - A_j^1 A_i^2 =-i2\; \frac{\( \pa_{i} u \pa_{j} u^* -
\pa_{j} u \pa_{i} u^*\)}{\( 1+\mid u \mid^2\)^2 }
\ee
and so
\be
\( A_i^1 A_j^2 - A_j^1 A_i^2\)^2 = 8\;
\frac{\mid \nabla u\mid^4 -
\( \nabla u\)^2\( \nabla u^*\)^2 }{\( 1+\mid u \mid^2\)^4 }
\ee
Consequently, the static energy reads
\br
E&=& \int d^3 x \; \(  \( A_j^1 A_j^1 + A_j^2 A_j^2\) +
 \( A_i^1 A_j^2 - A_j^1 A_i^2\)^2\)\nonu\\
&=& 16 \; \int d^3 x \; \(
\frac{\mid \nabla u \mid^2}{\( 1+\mid u \mid^2\)^2 } +
\frac{\mid \nabla u\mid^4 -
\( \nabla u\)^2\( \nabla u^*\)^2 }{\( 1+\mid u \mid^2\)^4 }\)
\er
Where the first line is eq. (12) of \ct{vB} (notice that  $e=1=m$ has been taken
  after a  $x\rightarrow x/em$  redefinition ).

One sees that as stated above, the first term invloves only the transverse colour components of the gauge field  and
 it formally coincides with the 
functional one chooses to minimize to get an  abelian gauge. This observation led  \ct{vB} to suggest  that  the minima of the
Skyrme-Faddeev, knot configurations with topological charge given by linking numbers, may
correspond to  the vacua of the nonabelian theory, fixed to pure gauge.

With our analysis one can obtain more precise information from the correspondence.  The second term, which respresents the
Hopf tensor, involves the diagonal components, as it is a commuatator in colour space, and so it does not corespond  strictly to the
maximal abelian gauge, where one only minimizes the transverse components.  Observe that, since
$A$ is flat, 
\be
A_i^1 A_j^2 - A_j^1 A_i^2 =-i(\pa_{i}A_j^{3}-\pa_{j}A_i^{3})
\ee
 and this term  is the curl of the third colour component, $\nabla \times {\bf A}^3$, so that those diagonal  degrees of freedom appear
as the dual (chromomagnetic) components. Both results are relevant for the claims that the minima of the Skyrme-Faddeev model may
represent QCD at long distances \ct{fadedual}. Moreover, this  Hopf term cannot
be directly proportional  \footnote {This simple argument is  what prevents Skyrme theory from saturating a BPS type of topological bound, which would require
proportionality, as explained in
\ct{Novikov}.} to  the Cartan component ${\bf A}^3$.
 Therefore, the static energy of Skyrme-Faddev model does not represent neither the functional corresponding to the space integral of
$A^2$, extensively studied in the context of non-perturbative QCD, including the flat case \ct{A2}, and which could provide information
about the sigma model solutions.

On the other hand, in the reduced sector, the second term in the energy has been simplified to $\mid \nabla u\mid^4 $,
since  $(\nabla u)^2$ vanishes due to the constraint. Therefore the energy  involves just the transverse $ {\bf A}^{\pm}$ components, and
so it is this sector, which  has a more simple expression, which represents 
the maximal abelian gauge functional.  Notice that from the point of view of the gauge theory, both from
the lattice and functional perturbative methods, there is no reason a priori for this vanishing contribution
$(\pa_{\mu} {\bf n}\cdot \pa_{\nu}{\bf n})^2$ to have the same coefficient as the remaining  $(\pa_{\mu} {\bf n})^4$  \ct{vW}. 
A comprehensive effort to find solutions,  which would improve
 the  physical analysis of the gauge fixed theory, is in progress.
One knows that the constraint is compatible with Derrick scaling and axial symmetry conditions \ct {Luiz}.
The alternative possible result that  
the integrable sector does not have solutions 
 would be also  relevant
for the connection with the gauge theory, as it would suggest that the strict maximal abelian gauge functional has
no stable minimum and   the effective
classical rotor spectrum.

We turn now to the possibility of adding a potential term as recently suggested \ct {fadedual} to break
the global colour symmetry and avoid massless modes and it can also be analysed in the integrability approach.
First, it is very difficult to maintain the $D_\mu {B}^\mu$
representation of the equation of motion, as one would have
to obtain a $V'$ term as the RHS of a commutator with $A_\mu$. Also, it
will be even more difficult to have
minimal energy configurations satisfying the constraint, which is
independent of the details of the potential.
But still, it is possible to have infinite conserved currents. For instance
if the potential depends just on
$\mid u^2 \mid$.
As an illustration one can consider the case required in $2+1$ dimensions
to stabilize the soliton and in
$3+1$ to account for the pion mass, given by  $V({\bf n})=\th (1+n_3)^4
={4 \th\over (1+|u|^2)^4}$.

In the restricted model, we have (using the equations of motion):
$$
D_\mu (B^{(j)})^\mu = {16\th\over (1+|u|^2)^4} (uP_{+1}-u^* P_{-1})
$$
for all values of $j$. If $\th=0$, then we have $2j+1$ conserved
currents for each $j$, using the usual procedure. But even for
$\th\not =0$ we still have one conserved current for each $j$. To
see this, define $J_\mu = g^{-1} (D_\mu B^\mu) g$, where $g$ is a generic
group element, and calculate
\br
\pa_\mu J^\mu &=& g^{-1} D_\mu (B^{(j)})^\mu g \nn
& = & {16 \th\over (1+|u|^2)^4} g^{-1}(uP_{+1}-u^*P_{-1})g
\er

For spin $j=1$ we can take a matrix-representation where
$$
uP_{+1}-u^*P_{-1} =
\left (\begin{array}{cc}0&u\\ u^*&0\end{array}\right )
$$
and it is easy to show that $uP_{+1}-u^*P_{-1}$ is invariant under
conjugation by $g$, and therefore $\pa^\mu J^{(1,0)}_\mu = 0$. For
higher spin, the calculation is a little more complicated, but also in
that case we find that $\pa^\mu J^{(j,0)}_\mu = 0$, thereby giving us
one conserved current for each value of $j$.

But on the other hand
it becomes more difficult that the constraint has any solutions, as
illustrated by the
$2+1$ dimensional case. There the restricted model is solved for the general
baby-skyrmion static solution given by  meromorphic fields $ u=\l (x+iy)$,
but  with the potential
considered, $u$ is a solution of the equations of motion  only for a
special value \ct{21} of the parameters
($\l = ({ {2  e}\th})^{1\over 4}$). Therefore, while it is still
possible to  have infinite conserved currents, the chances of an stable
solution are reduced considerably.

In conclusion, we have shown 
that there is some evidence for the Skyrme-Faddeev model representing global properties of the pure
 non-abelian theory in the infrared  and  capturing its topological properties in a local formulation,
which may be  solvable, but that some ingredients are still missing and more work
is required.
We have seen how this generalized zero curvature method can
be useful for this. The interplay between auxiliary gauge fields and gauge choices which contains, 
can be applied to other models. In particular  many expressions in this analysis  can be used 
directly to the Skyrme model in $2+1$ dimensions and  for generalized $CP^1$ models in any dimension which give useful non-perturbative
information about confinement.

\lskip{\bf Acknowledgments.} The author gratefully acknowledges IFT S\~ao Paulo, CERN and the University of Miami for hospitality
 and Orlando Alvarez,  M. Asorey, P.van Baal and M.Garcia P\'erez for discussions. 
Suggestions and comments of L.A. Ferreira were specially important. This work is supported by a
Spanish DGI  grant AEN99-0589.

\lskip
\end{document}